\documentclass[runningheads]{llncs}
\usepackage[T1]{fontenc}
\usepackage{graphicx,verbatim, caption}
\usepackage{booktabs}
\usepackage{comment}
\usepackage{subcaption}
\usepackage{hyperref}
\hypersetup{
    colorlinks=true,     
    urlcolor=blue,        
    citecolor=black,     
    linkcolor=black,     
}
\usepackage{color}

\usepackage{amsmath}
\usepackage{amssymb}

\begin{document}

\title{CALF: A Conditionally Adaptive Loss Function to Mitigate Class-Imbalanced Segmentation}
%

\author{Bashir Alam\inst{1} \and
Masa Cirkovic\inst{1} \and
Mete Harun Akcay\inst{1} \and
Md Kaf Shahrier\inst{1} \and
Sebastien Lafond\inst{1} \and
Hergys Rexha\inst{1} \and
Kurt Benke\inst{2} \and
Sepinoud Azimi\inst{3} \and
Janan Arslan\inst{4,5} 
}

\authorrunning{B. Alam et al.} 

\institute{
Åbo Akademi University, Faculty of Science and Engineering, Finland \\
\and
School of Engineering, University of Melbourne
Parkville, Victoria, Australia \\
\and
Institute for Health Systems Science, Delft University of Technology, The Netherlands \\
\and
Sorbonne Université, CNRS, INSERM, AP-HP, Inria, Paris Brain Institute (ICM), Paris, France \\
\and
Centre for Eye Research Australia, University of Melbourne, Royal Victorian Eye \& Ear Hospital, East Melbourne, Australia
}

\maketitle              
\begin{abstract}

Imbalanced datasets pose a considerable challenge in training deep learning (DL) models for medical diagnostics, particularly for segmentation tasks. Imbalance may be associated with annotation quality, limited annotated datasets, rare cases, or small-scale regions of interest (ROIs). These conditions adversely affect model training and performance, leading to segmentation boundaries which deviate from the true ROIs. Traditional loss functions, such as Binary Cross Entropy, replicate annotation biases and limit model generalization. We propose a novel, statistically driven, conditionally adaptive loss function (CALF) tailored to accommodate the conditions of imbalanced datasets in DL training. It employs a data-driven methodology by estimating imbalance severity using statistical methods of skewness and kurtosis, then applies an appropriate transformation to balance the training dataset while preserving data heterogeneity. This transformative approach integrates a multifaceted process, encompassing preprocessing, dataset filtering, and dynamic loss selection to achieve optimal outcomes. We benchmark our method against conventional loss functions using qualitative and quantitative evaluations. Experiments using large-scale open-source datasets (i.e., UPENN-GBM, UCSF, LGG, and BraTS) validate our approach, demonstrating substantial segmentation improvements. Code availability: \href{https://anonymous.4open.science/r/MICCAI-Submission-43F9/}{https://anonymous.4open.science/r/MICCAI-Submission-43F9/}.


\end{abstract}

\section{Introduction}

The application of artificial intelligence (AI) and deep learning (DL) methods has become a revelation in the high-risk and high-stakes world of medicine, particularly in tasks such as image segmentation \cite{Litjens2017,Shen2017,Ronneberger2015}. Segmentation algorithms play a crucial role in isolating regions of interest (ROIs) from medical images, enabling disease diagnosis and biomarker discovery \cite{Sudre2017}. However, the clinical adoption of these methods faces critical challenges, notably the availability of large and high quality data sufficient for training \cite{Kamnitsas2017,Ma2021}. Furthermore, available annotations may be subject to several flaws, resulting in sparse annotations or those that fail to accurately capture true ROIs. These challenges contribute to data imbalance and compromise DL training and generalization \cite{Jadon2020}.

Loss function selection is crucial, as it plays a fundamental role in iteratively guiding the optimization of model parameters \cite{Ma2021,Yeung2021}. Conventional loss functions such as Binary Cross Entropy (BCE), Focal, Tversky, and Dice Losses have been widely adopted, some including characteristics that could mitigate class imbalance \cite{Sudre2017,Abraham2019}. However, they often suffer from inherent limitations that hinder their effectiveness, one of which includes their sensitivity to the quality and format of annotation \cite{Jadon2020}. Since most DL models learn based on the provided annotations, these loss functions tend to reinforce annotation biases. For example, if training labels are polygon-based, roughly drawn, or exceed the boundary limits of the true ROI, the model predictions are likely to mimic these idiosyncrasies, restricting generalization to complex-shaped ROIs (e.g., disease-affected regions with irregular or amorphous boundaries) \cite{Ma2021}. Traditional loss functions also exhibit difficulties in handling foreground-background imbalances where the ROI occupies a small fraction of the overall image. BCE overemphasizes the dominant background class and poorly segments small or low-contrast structures
\cite{Kamnitsas2017}. Focal Loss partially addresses this by down-weighting easy-to-classify pixels but requires careful hyperparameter tuning \cite{Abraham2019}. Dice and Tversky Losses consider pixel-wise agreement between predictions and ground truth, but are susceptible to over-segmentation and fail where under-segmentation is more detrimental \cite{Sudre2017}.These loss functions lack adaptability to varying dataset characteristics, sometimes requiring manual tuning \cite{Jadon2020}. The rigid nature of these functions limits their applicability to diverse segmentation tasks, necessitating the development of a more dynamic approach which can adjust to data heterogeneity and uncertainty.

In this paper, we introduce the \textit{Conditionally Adaptive Loss Function}(CALF), a novel, statistically driven approach for segmentation tasks on imbalanced datasets. Our key contributions include: (1) developing a dynamic loss function that adapts to the statistical characteristics of the dataset, mitigating annotation biases and enhancing segmentation performance; (2) a hybrid data processing approach integrates preprocessing techniques, flexible dataset filtering mechanisms, and dynamic loss function selection to address data scarcity and imbalance while preserving dataset heterogeneity; (3) a configurable data filtering system that introduces a dataset balancing mechanism, allowing controlled variation in the ratio of ROI-present to ROI-absent images, thereby enabling robust model evaluation under different data conditions; and (4) a comprehensive performance evaluation in which we compare CALF against various loss functions in tumor segmentation tasks, demonstrating superior generalization in rare-class segmentation scenarios.

\section{Methodology}
 Let \( \mathbf{x}_i \in \mathbb{R}^{w \times h} \) represent an input grayscale image, where \( w \) and \( h \) denote the width and height of the image, respectively. Each image is associated with a binary segmentation mask \( \mathbf{y}_i \in \{0,1\}^{w \times h} \), where the pixel values indicate the foreground and background regions. Given a dataset of images \( \{ \mathbf{x}_1, \dots, \mathbf{x}_N \} \) and their corresponding masks \( \{ \mathbf{y}_1, \dots, \mathbf{y}_N \} \), the objective is to train a model \( f_{\mathbf{w}} \) with parameters \( \mathbf{w} \) that accurately predicts the segmentation mask for any given input image. Loss functions guide model optimization to distinguish between foreground and background regions. However, conventional loss functions do not account for statistical distributions of segmented objects across different images. This hinders model performance, as real-world medical datasets often exhibit uneven distributions of foreground regions. Therefore, the proposed approach dynamically adjusts according to skewness and kurtosis measures to analyze the distribution of foreground regions and counters with appropriate transformations.

\subsection{Skewness and Kurtosis}
Skewness quantifies the asymmetry of a probability distribution, which means it describes the distribution of foreground object sizes within a dataset. Kurtosis describes the shape of a probability distribution by measuring whether it is peaked or flat, reflected by the variability of the foreground object size. The definitions are described in Table~\ref{table:skewkurt} and include the foreground areas \( A = \{A_1, A_2, \dots, A_N\} \), the mean \( \mu \) and the standard deviation \( \sigma \). \( S < 0 \) indicates a distribution with a longer left tail (that is, larger objects in the foreground are more common), \( S > 0 \) implies the distribution has a longer right tail (i.e., smaller objects dominate), and \( S \approx 0 \) suggests the sizes of the foreground objects are more symmetrically distributed across the dataset. A high kurtosis (\( K > 0 \)) indicates a distribution with a sharp peak and heavy tails (that is, a mix of very large and small objects). A low kurtosis (\( K < 0 \)) corresponds to a flatter distribution, suggesting that the size of the foreground objects is more uniform across the dataset. Thresholds describing skewness and kurtosis level have been long-established \cite{Bulmer1979}.

\begin{table}[htbp]
\centering
\caption{Statistical moments defined as mean, standard deviation, skewness and kurtosis. CALF utilizes skewness and kurtosis to identify imbalances in the raw data distribution that could impact DL training.}
\begin{tabular}{l l l} 
\toprule
\textbf{Name} & \textbf{Indicator} & \textbf{Formula} \\
\midrule

Mean
  & Central Tendency
  & 
    \(\displaystyle \mu = \frac{1}{N} \sum_{i=1}^{N} A_i \quad
    \)
  \\
  
Standard Deviation
  & Dispersion
  & 
    \(\displaystyle \sigma = \sqrt{\frac{1}{N} \sum_{i=1}^{N} (A_i - \mu)^2}
    \)
  \\

Skewness
  & Asymmetry
  & 
    \(\displaystyle S = \frac{ \frac{1}{N} \sum_{i=1}^{N} (A_i - \mu)^3 }{ \sigma^3 }    \)
  \\

Kurtosis
  & Tailedness/Peakness
  & 
    \(\displaystyle K = \frac{ \frac{1}{N} \sum_{i=1}^{N} (A_i - \mu)^4 }{ \sigma^4 } - 3
    \)
  \\

\bottomrule
\end{tabular}
\label{table:skewkurt}
\end{table}

\subsection{Conditionally Adaptive Loss Function}
The proposed loss function, CALF,  can be formulated as described in Table~\ref{table:lossfunctions}. The skewness and kurtosis describe the raw data distribution (i.e., whether the distribution of values is extreme or moderate), which are then used as indicators in identifying the most appropriate transformation. These transformations are designed to minimize and stabilize variance and improve normality by handling small and large values as shown below \cite{Bulmer1979,Stevens2009}:
\begin{itemize}

    \item \(S \leq -1\): Fisher transformation compresses the large objects found. 
    \item \(-1 < S \leq -0.5\): shows large objects present but not dominant. Logit slightly expands small and compresses large values.
    \item \(-0.5 < S < 0\): indicates sparse bright regions. Arcsine expands extreme values while compressing mid-range values.
    \item \(S \geq 1\): Log10 improves separation of small foreground objects.
    \item \(0.5 \leq S < 1\): involves small objects but not dominant. Natural log moderately expands small and compresses large objects.
    \item \(0 < S \leq 0.5, K < 0\): show a uniform distribution with Log10 simply spreading out smaller regions for better separation. 
    \item \(0 < S \leq 0.5, K \geq 0\): has many small but not extreme objects, with Log10 balancing size and distribution. 
\end{itemize}

The loss function dynamically adapts to the various distributional properties of any given dataset, ensuring that segmentation models remain sensitive to variations in the size of objects and their distribution. The method stabilizes optimization, prevents bias towards dominant object sizes, and improves segmentation performance across diverse datasets.

\begin{table}[htbp]
\centering
\caption{Conditionally adaptive loss function. The loss function automatically detects raw data distribution (i.e., skewness and kurtosis) and applies the appropriate transformation to counteract the imbalance. Unlike other loss functions, such as Focal or Tversky, this method requires no user input.}
\begin{tabular}{l l p{7cm}} 
\toprule
\textbf{Transformation} & \textbf{Condition} & \textbf{Formula} \\
\midrule
\(\displaystyle \mathcal{L}_{\text{Fisher}}\)
  & \(\displaystyle S \le -1\)
  & 
    \(\displaystyle - \mathbb{E} \Bigl[
     y \cdot \tfrac{1}{2} \ln\Bigl(\tfrac{1 + p}{1 - p}\Bigr)
     + (1 - y) \cdot \tfrac{1}{2} \ln\Bigl(\tfrac{1 + (1 - p)}{1 - (1 - p)}\Bigr)
     \Bigr]
    \)
  \\
\(\displaystyle \mathcal{L}_{\text{Logit}}\)
  & \(\displaystyle -1 < S \leq -0.5 \)
  & 
    \(\displaystyle - \mathbb{E} \Bigl[ y \cdot \ln \frac{p}{1 - p} + (1 - y) \cdot \ln \frac{1 - p}{p} \Bigr]
    \)
  \\

\(\displaystyle \mathcal{L}_{\text{Arcsine}}\)
  & \(\displaystyle -0.5 < S < 0 \)
  & 
    \(\displaystyle - \mathbb{E} \Bigl[ y \cdot \arcsin(\sqrt{p}) + (1 - y) \cdot \arcsin(\sqrt{1 - p}) \Bigr]
    \)
  \\

\(\displaystyle \mathcal{L}_{\text{Log10}}\)
  & \(\displaystyle S \geq 1 \)
  & 
    \(\displaystyle - \mathbb{E} \Bigl[ y \cdot \log_{10}(p) + (1 - y) \cdot \log_{10}(1 - p) \Bigr]
    \)
  \\

\(\displaystyle \mathcal{L}_{\text{Natural Log}}\)
  & \(\displaystyle  0.5 \leq S < 1  \)
  & 
    \(\displaystyle - \mathbb{E} \Bigl[ y \cdot \ln p + (1 - y) \cdot \ln (1 - p) \Bigr]
    \)
  \\

\(\displaystyle \mathcal{L}_{\text{Log10}}\)
  & \(\displaystyle 
  \begin{aligned}
  0 < S \leq 0.5 \\ 
  \text{and } K < 0
  \end{aligned}
  \)
  & 
    \(\displaystyle - \mathbb{E} \Bigl[ y \cdot \log_{10}(p) + (1 - y) \cdot \log_{10}(1 - p) \Bigr]
    \)
  \\

\(\displaystyle \mathcal{L}_{\text{BCE-Dice}}\)
  & \(\displaystyle 
  \begin{aligned}
    0 < S \leq 0.5\\
    \text{and } K \geq 0 
      
  \end{aligned} \)
  & 
    \(\displaystyle \mathcal{L}_{\text{Dice}} = -\frac{2 \sum (y p) + \epsilon}{\sum y + \sum p + \epsilon} \text{ ;  } \mathcal{L}_{\text{BCE}} + (1 - \mathcal{L}_{\text{Dice}})
    \)
  \\

\bottomrule
\end{tabular}
\label{table:lossfunctions}
\end{table}

\section{Experiments}

\begin{figure}
\includegraphics[width=\textwidth]{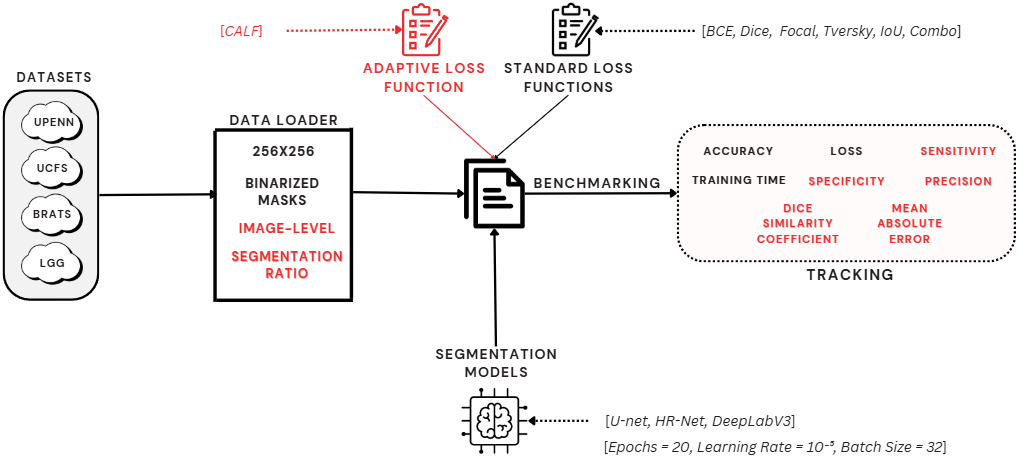}
\caption{Workflow of the experiments.} 
\label{fig:workflow}
\end{figure}

CALF performance was evaluated on the basis of the workflow outlined in Figure~\ref{fig:workflow}. It involves the extraction of four open source datasets (Sec.~\ref{section:datasets}), preprocessing using a custom data loader, benchmarking which involves comparing CALF against selected loss functions and segmentation models, and tracking both qualitative and quantitative performance (Sec.~\ref{section:benchmarking} and Sec.~\ref{section:results}).

\subsection{Datasets}
\label{section:datasets}
We conducted experiments using four open source high-quality brain cancer datasets from The Cancer Imaging Archive (TCIA) \cite{TCIA2025} (Table~\ref{tab:dataset_split_summary}). These were chosen to capture various imbalanced conditions for testing and validating CALF. The \textbf{UCSF-PDGM} \cite{UCSF-PDGM}, \textbf{BraTS} \cite{BraTS},  \textbf{UPENN-GBM} \cite{UPENN-GBM}, and \textbf{LGG-1p19q\\Deletion} \cite{LGG-1p19qDeletion} datasets contain gliomas and glioblastomas MRI scans. 

The datasets contain combinations of T1, T2, T1 contrast-enhanced (T1 CE), FLAIR, DWI and SWI. Resolutions ranged from 240×240 to 256×256 pixels. 
The data processing involved the conversion of three-dimensional images into two-dimensional slices, which were subsequently saved in the Portable Network Graphics (PNG) format. The ground truth labels were converted from grayscale to binary in all datasets. The final dataset consisted of 1,410 patients and 589,838 2D images. An overview of the data (including the training and testing divisions) is given in Table~\ref{tab:dataset_split_summary}.

\begin{table}[h]
    \centering
    \caption{Summary of dataset used in experiments, including total patient and image numbers, as well as division into training and testing.}
    \setlength{\tabcolsep}{3pt}
    \begin{tabular}{lcccc}
        \hline
        \textbf{Dataset} & \textbf{Patients} & \textbf{Images} & \textbf{Training Images} & \textbf{Testing Images} \\
        \hline
        UPENN-GBM & 611 &  284,115 & 256,620 & 27,495 \\
        UCSF & 495 & 230,175 & 207,900 & 22,275 \\
        BraTS & 145 & 67,425 & 60,900 & 6,525 \\
        LGG & 159 & 8,123 & 7,328 & 795 \\
        \hline
        \textbf{Total} & 1,408 & 589,838 & 532,748 & 57,090 \\
        \hline
    \end{tabular}
    \label{tab:dataset_split_summary}
\end{table}
\raggedbottom

\subsection{Benchmarking}
\label{section:benchmarking}

We trained our datasets using U-Net \cite{Ronneberger2015}, DeepLabV3 \cite{Chen2017}, and FPN \cite{Kirillov2019} and seven different loss functions (BCE \cite{Goodfellow2016}, Dice \cite{Milletari2016}, Tversky \cite{Salehi2017}, IoU \cite{Rahman2016}, Focal \cite{Lin2017}, BCE-Dice \cite{Isensee2021}, and CALF). Comparative loss functions were selected after reviewing 25 different loss functions identified by querying the \texttt{\{loss\_function
\\\_name\} segmentation} in the \href{https://www.dimensions.ai/}{Dimensions.ai} database. These were sorted based on: citation count, use cases (particularly for imbalanced data), and their frequent application in segmentation tasks. A custom data loader was created to specify a tumor-to-non-tumor ratio (from 0 to 1), ensuring models were provided with annotated image–mask pairs, as imbalances also included an inequitable distribution of images with tumors vs. those without. 
The quantitative metrics used are shown in Figure \ref{fig:workflow}. These were collected along with qualitative analysis that evaluated the precision of the segmentation through visual inspection. 

\section{Results}
\label{section:results}
\begin{figure}[!htbp]
    \centering

    \begin{subfigure}{\textwidth}
        \centering
        \includegraphics[width=0.88\textwidth]{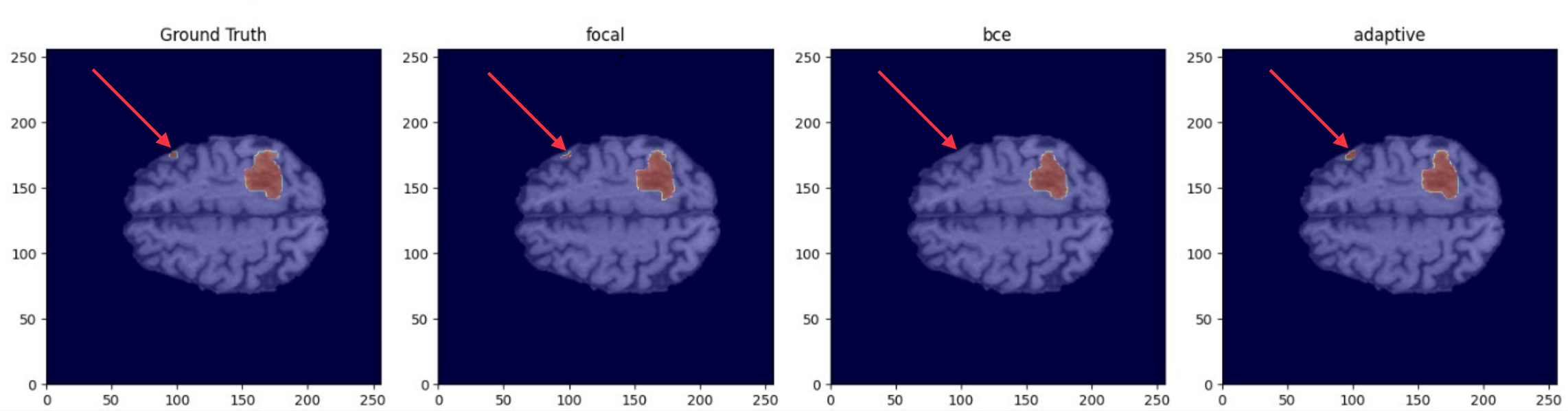}
        \caption{\scriptsize CALF captures small regions (red arrow) not detected by other loss functions.}
        \label{fig:comparison}
    \end{subfigure}

    \begin{subfigure}{\textwidth}
        \centering
        \includegraphics[width=0.65\textwidth]{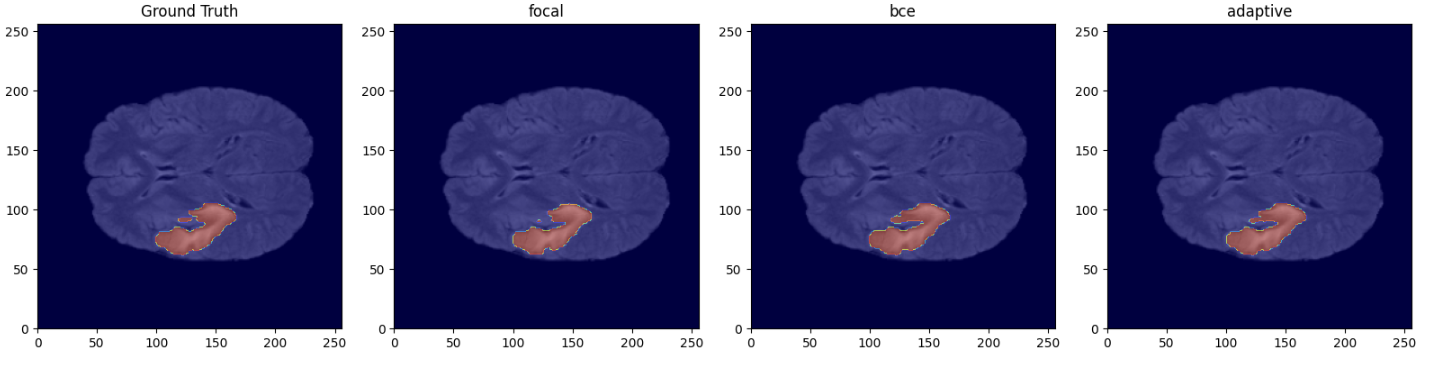}
        \includegraphics[width=0.65\textwidth]{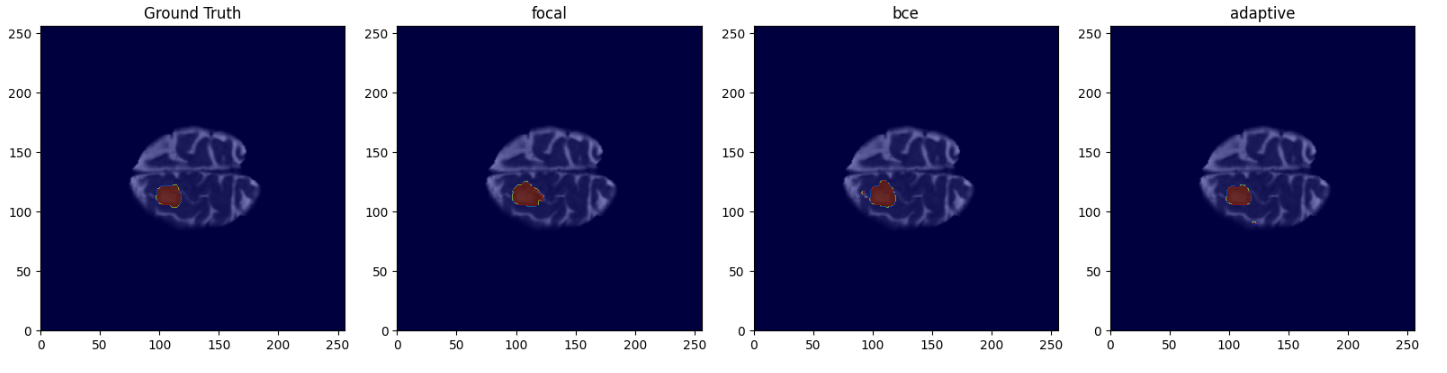}
        \includegraphics[width=0.65\textwidth]{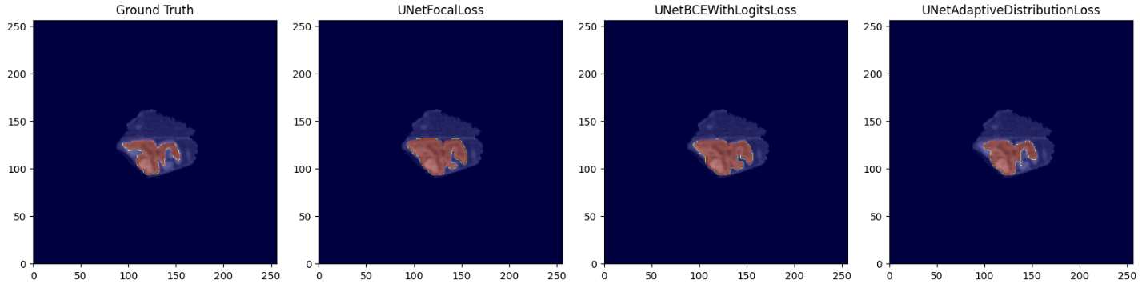}
        \caption{\scriptsize CALF more precisely captures the tumor region boundary compared to Focal Loss and BCE, which typically over-segment.}
        \label{fig:combined_precise}
    \end{subfigure}

    \begin{subfigure}{\textwidth}
        \centering
        \includegraphics[width=0.65\textwidth]{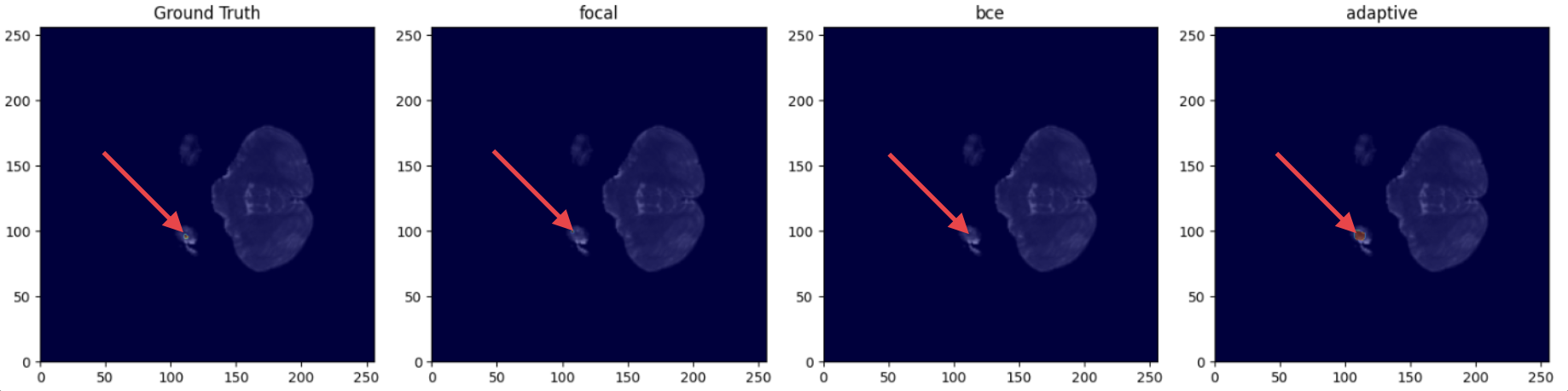}
        \includegraphics[width=0.65\textwidth]{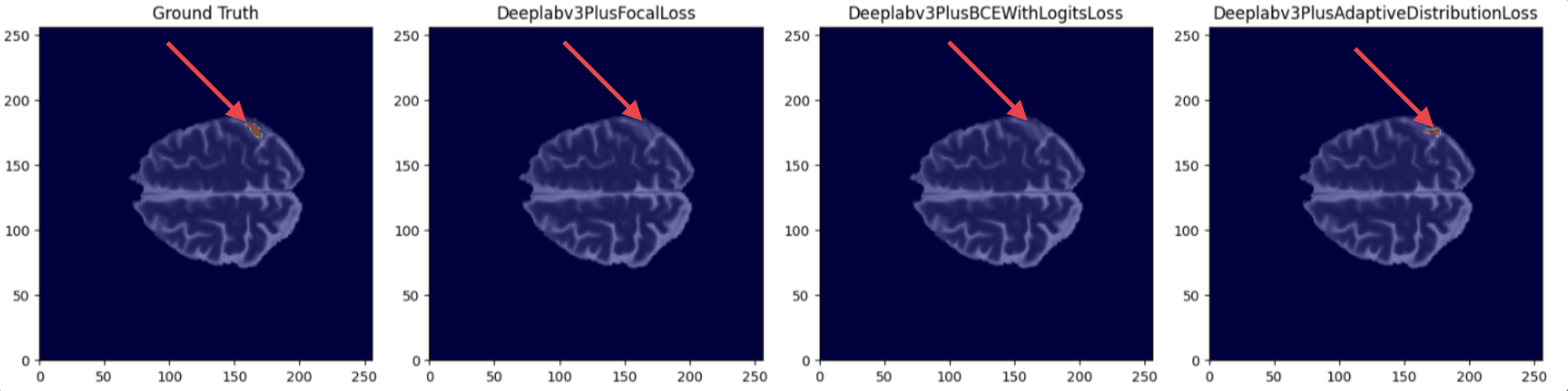}
        \caption{\scriptsize Examples where CALF successfully captured very small tumor regions, while BCE and Focal failed.}
        \label{fig:combined_only}
    \end{subfigure}
    
    \begin{subfigure}{\textwidth}
        \centering
        \includegraphics[width=0.88\textwidth]{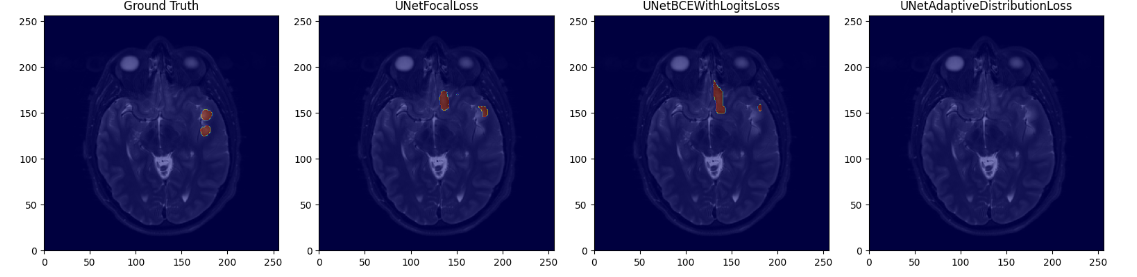}
        \caption{\scriptsize An instance where CALF did not predict a mask that was captured by Focal loss and BCE.} 
        \label{fig:unet-min-dsc-bce}
    \end{subfigure}
    
    \caption{Comparison of different loss functions' performance.}
    \label{fig:combined_all}
\end{figure}

Several combinations of model-loss functions with varying ratios were tested to determine how adaptable the loss functions are under various data scarcity conditions. Table~\ref{tab:model_comparison} presents a comparative analysis of three best performing loss functions (BCE, Focal, and CALF) across the models, with a ratio of \textbf{40.9\%} (the `default' ratio, representing the total number of tumor cases available). Our proposed loss function CALF demonstrated consistent performance, quantitatively competing with BCE and outperforming Focal loss in multiple cases. Figure \ref{fig:combined_all} also demonstrates CALF segmentation performance in comparison with BCE and Focal Loss.

    



\begin{table*}[htbp]
\centering
\caption{Model Performance with BCE, Focal, and CALF with ratio of 0.409.}
\begin{tabular}{c|ccc|ccc|ccc}
\hline
\multicolumn{1}{c}{\textbf{Model}} & \multicolumn{3}{c}{\textbf{BCE}} & \multicolumn{3}{c}{\textbf{Focal}} & \multicolumn{3}{c}{\textbf{CALF}} \\
  & Accuracy & DSC & MAE & Accuracy & DSC & MAE & Accuracy & DSC & MAE \\\hline

U-Net & \textbf{0.9991} & \textbf{0.9589} & \textbf{0.0008} & 0.9990 & 0.9552 & 0.0009 & 0.9990 & 0.9577 & 0.0009 \\

\begin{tabular}{@{}c@{}}DeepLab \\ v3\end{tabular} & \textbf{0.9969} & 0.8597 & \textbf{0.003} & 0.9961 & 0.7998 & 0.0037 & \textbf{0.9969} & \textbf{0.8598} & \textbf{0.003} \\

FPN & \textbf{0.9981} & \textbf{0.9128} & \textbf{0.0018} & 0.9977 & 0.8892 & 0.0022 & 0.9979 & 0.9072 & 0.002 \\\hline

\end{tabular}
\label{tab:model_comparison}
\end{table*}

As evident in Table \ref{tab:model_comparison}, BCE showed strong overall performance, achieving the highest DSC and accuracy when trained on U-Net, while the proposed CALF performed equally well and even surpassed BCE in DeepLabV3. The Focal loss function, on the other hand, struggled significantly in DeepLabV3, with a DSC score of only 0.7998, highlighting its weakness in this architecture.

To further investigate the performance of CALF, we trained FPN on an imbalanced dataset with a ratio of \textbf{10\%}. The outcomes presented in Table~\ref{tab:imbalanced_dataset}, demonstrate that the proposed CALF outperformed alternative loss functions in this particular scenario, which is the focal point of our investigation. Table~\ref{tab:imbalanced_dataset} highlights the robustness of CALF, which consistently outperformed other loss functions in the case of imbalanced dataset. 
This finding underscores the efficacy of our loss function in addressing severe class imbalance. Although the BCE loss function exhibited specificity, it demonstrated challenges in detecting small tumor regions. In contrast, CALF exhibited more consistent performance in this regard.

\begin{table}[h]
\centering
\footnotesize
\caption{FPN model performance for various loss functions with a ratio of 0.1.}
\begin{tabular}{lccccccc}
\hline
\textbf{Loss} & \textbf{Accuracy} & \textbf{DSC} & \textbf{Specificity} & \textbf{Sensitivity} & \textbf{Precision} & \textbf{MAE} \\\hline
BCE & 0.9964 & 0.8187 & 0.9992 & 0.7430 & 0.9174 & 0.0035 \\
Tversky & 0.9943 & 0.7548 & 0.9964 & 0.8135 & 0.7106 & 0.0056 \\
IoU & 0.9945 & 0.7683 & 0.9962 & 0.8473 & 0.7092 & 0.0054 \\
Focal & 0.9957 & 0.7701 & \textbf{0.9996} & 0.6488 & \textbf{0.9573} & 0.0042 \\
Dice & 0.9948 & 0.7722 & 0.9967 & 0.8218 & 0.7343 & 0.0051 \\
BCE-Dice & 0.9961 & 0.8265 & 0.9976 & \textbf{0.8534} & 0.8046 & 0.0038 \\
CALF & \textbf{0.9965} & \textbf{0.8267} & 0.9991 & 0.7598 & 0.9113 & \textbf{0.0034} \\\hline
\end{tabular}
\label{tab:imbalanced_dataset}
\end{table}
\raggedbottom

\section{Conclusion}

CALF, a conditionally adaptive loss function, was introduced to address the challenges posed by class imbalance in medical image segmentation. By leveraging statistical characteristics like skewness and kurtosis, appropriate transformations are applied to mitigate imbalance and optimize learning. Experiments were conducted on four large-scale, open-source tumor segmentation datasets: UCSF-PDGM \cite{UCSF-PDGM}, BraTS \cite{BraTS}, UPENN-GBM \cite{UPENN-GBM} and LGG \cite{LGG-1p19qDeletion}. CALF exhibited a consistent improvement in segmentation accuracy, achieving high-level and persistent qualitative and quantitative results compared to standard loss functions, which exhibited variability in their performance. For example, BCE quantitatively performed well, particularly in its detection of larger tumor regions. However, qualitatively, it over-segmented ROIs, illustrating poor performance in rare-class scenarios. While CALF provides a promising approach to improving medical image segmentation, future research could also explore its adaptability to varying noise levels and integration with semi-supervised learning techniques. Additionally, expanding its application to other imaging modalities and segmentation tasks could further demonstrate its flexibility and validate its robustness. 

\end{document}